# Dispersive coupling and optimization of femtogram L3-nanobeam optomechanical cavities

Jiangjun Zheng,[1,3,*] Xiankai Sun,[2,3] Ying Li,[1] Menno Poot,[2] Ali Dadgar,[1] Hong X. Tang,[2] and Chee Wei Wong[1,*]

[1] *Optical Nanostructures Laboratory, Columbia University, New York, New York 10027, USA*
[2] *Department of Electrical Engineering, Yale University, New Haven, Connecticut 06511, USA*
[3] *These authors contributed equally to this work*
[*]*jz2356@columbia.edu, cww2104@columbia.edu*

**Abstract:** We present the design of a femtogram L3-nanobeam photonic crystal cavity for optomechanical studies. Two symmetric nanobeams are created by placing three air slots in a silicon photonic crystal slab where three holes are removed. The optical quality factor ($Q$) is optimized up to 52,000. The nanobeams' mechanical frequencies are higher than 600 MHz due to their femtogram effective modal masses. The optical and mechanical modes are dispersively coupled with a vacuum optomechanical coupling rate $g_0/2\pi$ exceeding 200 kHz. The anchor-loss-limited mechanical $Q$ of the differential beam mode is evaluated to be greater than 10,000 for structures with ideally symmetric beams. The influence of variations on the air slot width and position is also investigated. The devices can be used as ultrasensitive sensors of mass, force, and displacement.

**OCIS codes:** (220.4880) Optomechanics; (230.5750) Resonators; (230.5298) Photonic crystals.

## References and links

## 1. Introduction

Cavity optomechanics, a subject studying the coherent interaction of optical and mechanical degrees of freedom of various optical cavities, has been a recent research focus [1−4]. The topics include laser cooling of mesoscopic systems to their motional quantum mechanical ground state [5−13], photon–phonon and acoustic transduction and storage of light pulses [14−16], and quantum precision measurements of microwave and optical photons [17−19]. Photonic crystals (PhC) are a versatile cavity platform, which has been widely used for light–matter and light–structure interactions, e.g., cavity quantum electrodynamics [20,21], nonlinear optics [22–24], and cavity optomechanics [25,26]. By taking advantage of optical gradient forces [27–29], high-$Q$ PhC cavities exhibit strong optomechanical interactions in both one-dimensional (1D) [30] and two-dimensional (2D) [31] geometries.

In this paper we present the details of theoretical modeling and design of a "nanobeam-in-cavity" we experimentally demonstrated recently [32], where the nanobeams are embedded *within* a small PhC cavity to obtain a small modal mass and large optomechanical coupling. The mechanical modes are localized at smaller length scales than the optical modes, with femtogram masses. The mechanical properties are readily and deterministically tuned. The strong optical scattering from the embedded nanobeams requires reengineering of the photonic band structure and fine tuning of the radiation light cone to optimize the optical $Q$. We simulate the mechanical $Q$ and investigate the influence of fabrication imperfections [33,34]. We further analyze the optomechanical coupling rate $g_{om}/2\pi$ for several nanobeam-in-cavity geometries, and obtain a value up to 15.5 GHz/nm (corresponding to a vacuum optomechanical coupling rate $g_0/2\pi$ of 326.6 kHz).

## 2. Optical design: band structure and radiation suppression

Nanobeams are widely used to build high-frequency nanomechanical resonators [35−37]. As shown in Fig. 1, our nanobeam-in-cavity consists of two nanobeams embedded in a PhC slab where three air slots are placed in the region of three missing holes of a triangular lattice. Since a PhC cavity based on three missing air holes in an otherwise perfect triangular lattice is usually referred to as an L3 cavity [38,39], we name ours an "L3-nanobeam cavity." The nanobeams introduce strong perturbation to the original L3 cavity, resulting in significant modification of the optical characteristics [40]: First, the effective refractive index in the cavity region is reduced by the slots, which shifts the cavity resonance away from the bandgap and makes it difficult to localize the optical energy in a small volume; Second, the sharp edges of the slots result in a large radiation energy leakage [38]. Careful designs are thus required to restore a high optical $Q$ after introducing the slots. We overcome the adverse effects and optimize the cavity with the following strategies: First, the $y$ distance of $w_{wg}$ is increased, i.e., made wider than the width of a single-missing-row (W1) waveguide, which compensates the effect of reduced effective refractive index. The holes in the cavity row are enlarged accordingly to provide a reliable in-plane field confinement in the $x$ direction. The holes surrounding the cavity are shifted to tune the optical field profile such that the vertical radiation scattering is minimized [38]. Numerical modeling proved the effectiveness of the above implementations and that the optical field of this L3-nanobeam cavity is indeed different from that of a regular L3 cavity [38]. For simplicity, it is straightforward to construct symmetric structures with $w_b = w_{b1} = w_{b2}$ and $w_{as} = w_{as1} = w_{as2} = w_{as3}$.

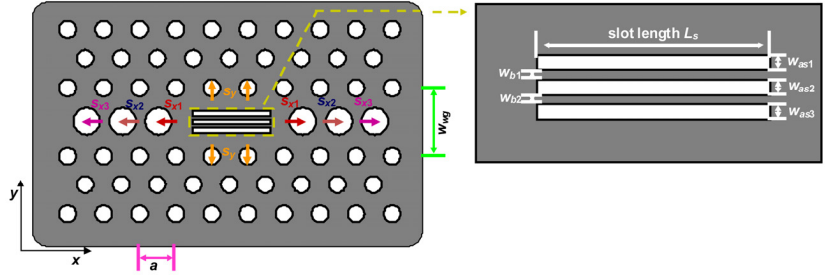

Fig. 1. Geometry of the L3-nanobeam cavity. Left: overview of the device. Right: zoom-in of the beam region. $a$ is the lattice constant of the triangular PhC; $s_{x1}$, $s_{x2}$, and $s_{x3}$ are the hole offsets in the $x$ direction; $s_y$ is the hole offset in the $y$ direction; $w_{b1}$ and $w_{b2}$ are beam widths; $w_{as1}$, $w_{as2}$, and $w_{as3}$ are slot widths.

The intrinsic properties of different regions of the L3-nanobeam structures are studied by band structure analysis with MPB [41], a vectorial eigensolver of Maxwell's equations with periodic boundary conditions. It is well known that the localized modes of a regular L3 cavity can be viewed as Fabry–Pérot modes from a guided TE-like $y$-odd band of a W1 waveguide. Similarly, the optical modes of the L3-nanobeam cavity are also based on a slotted waveguide band. Fig. 2(a) shows the band structure of the TE-like modes for the PhC lattice, where a quasicomplete bandgap covers the C band and provides confinement for the optical field in the $y$ direction. Fig. 2(b) shows the band structure of the slotted waveguide for the TE-like $y$-odd waveguide modes, where solid and dashed lines correspond to $(w_{as}, w_b) = (60\text{ nm}, 80\text{ nm})$ and $(w_{as}, w_b) = (60\text{ nm}, 60\text{ nm})$, respectively. The waveguide width $w_{wg}$ is selected to be $1.35 \times \sqrt{3}a$, with which value the slot waveguide band is well located at the center of the PhC bandgap. Fig. 2(c) shows the mode-edge frequency versus the waveguide width $w_{wg}$. The gray regions indicate the slab mode continua of the PhC lattice. A regular W1 waveguide has a propagating TE-like $y$-odd band inside the PhC bandgap; however, when the slots are introduced, the slotted W1 waveguide band shifts up, pushing its bandedge into the upper PhC slab continuum. By increasing $w_{wg}$ from $\sqrt{3}a$ to $1.35 \times \sqrt{3}a$, the waveguide band moves back to the center of the bandgap and the optical field is guided again by the nanobeam PhC waveguide [Fig. 2(d) and (e)]. The $E_y$ field in the three slots is strong, similar to that of other air-slot PhC waveguides [42]. The intensity $|E_y|^2$ across the waveguide is shown in Fig. 2(f), where the origin ($y = 0$) denotes the center of the middle slot. The boundary conditions result in sharp slopes at the air–silicon interfaces. The fields are highly concentrated in the slots, as indicated by the three peaks. The center peak is weaker than the outer ones, providing asymmetric field across the silicon nanobeams, a prerequisite for taking advantage of the gradient optical force [27]. Next, to form a cavity mode, the field has to be confined in the $x$ direction. Therefore, the infinitely long slots are replaced with ones of the planned beam length, surrounded by "mirror" air holes, to create a cavity [43]. Fig. 3(a) shows the TE-like $y$-odd guided band for the "mirror" waveguide with hole radius $r_{wg}$ = 160 nm. Fig. 3(b) shows that the edge of this guided band is pushed up by using a larger radius. Concluded from a comparison between Fig. 3(a) and Fig. 2(b), the band of the "mirror" waveguide sits well above that of the slot waveguide. Their zero overlap in frequency range is important for suppressing the optical energy leakage via the "mirror" waveguide. Based on the above guidelines, a selected set of geometrical parameters are: $(a, r, h, w_{wg}, r_{wg})$ = (430 nm, 0.29$a$, 220 nm, W1.35, 160 nm), which are used for further optimization of the cavity mode.

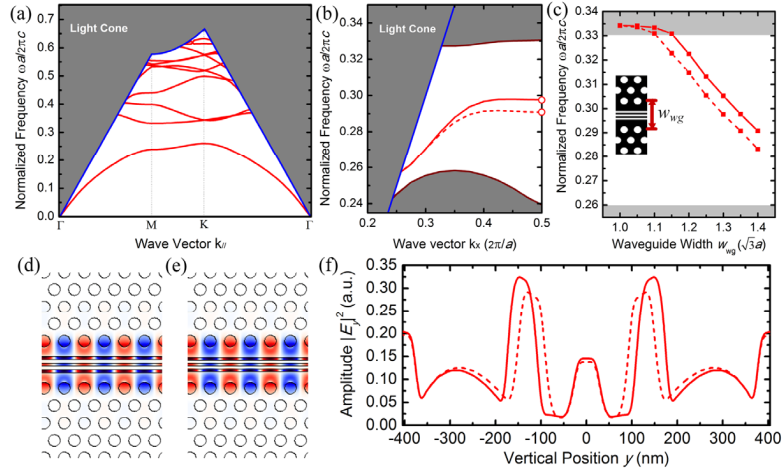

Fig. 2. (a) TE-like bands of a PhC slab with a triangular lattice of air holes with lattice constant $a$ = 430 nm and hole radius $r$ = 0.29$a$, refractive index of silicon $n_{si}$ = 3.48, and silicon slab thickness $h$ = 220 nm. (b) TE-like $y$-odd waveguide band for the slotted W1.35 waveguide with $w_{wg}$ = 1.35 × $\sqrt{3}a$, $(w_{as}, w_b)$ = (60 nm, 80 nm) (red solid line) and $(w_{as}, w_b)$ = (60 nm, 60 nm) (red dashed line). Red circles indicate the bandedges. (c) Waveguide bandedge frequency versus the waveguide width $w_{wg}$. The gray regions indicate the PhC slab mode continua. The inset illustrates the geometry. (d),(e) Field distribution of $E_y$ at the bandedge as indicated by the upper and lower red circles in (b), respectively. (f) $E_y$ intensity profile along the $y$ direction, i.e., perpendicular to the slots. The solid line is obtained from (d) and the dashed from (e), cut from center of the hot optical spots.

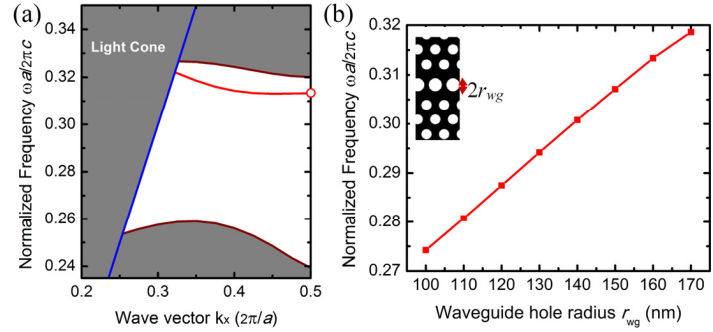

Fig. 3. (a) Band structure of the TE-like $y$-odd band for the "mirror" waveguide with hole radius $r_{wg}$ = 160 nm. (b) Waveguide bandedge frequency versus the waveguide hole radius $r_{wg}$. The inset illustrates the geometry.

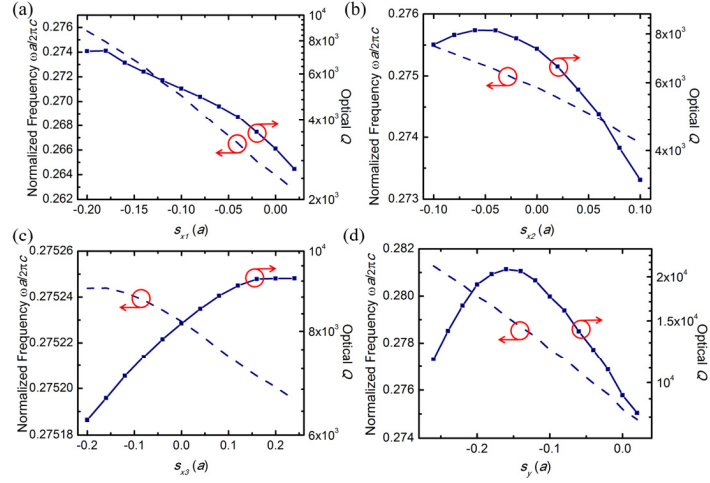

Fig. 4. Optimization process for the cavity with ($w_{as}$, $w_b$) = (60 nm, 80 nm) by tuning hole positions of $s_{x1}$, $s_{x2}$, $s_{x3}$, and $s_y$, respectively, in series. For an air slot length $L_s$ of 1.8$a$, the maximum optical $Q$ of 1.95 × 10$^4$ is achieved with ($s_{x1}$, $s_{x2}$, $s_{x3}$, $s_y$) = (−0.18$a$, −0.06$a$, 0.22$a$, −0.15$a$) [Design 1]. With a similar optimization process, cavities with ($w_{as}$, $w_b$) = (60 nm, 60 nm) achieve a higher optical $Q$ of 5.22 × 10$^4$ with ($L_s$, $s_{x1}$, $s_{x2}$, $s_{x3}$, $s_y$) = (1.9$a$, −0.3$a$, −0.02$a$, 0.1$a$, 0) [Design 2]. Solid lines indicate the optical $Q$, while the dashed ones indicate the normalized resonant frequency.

With band structure calculations, we have created a localized cavity made by ensuring the in-plane modal confinement. In what follows we use finite-difference time-domain (FDTD) method to simulate the L3-nanobeam cavities [44] and optimize their optical $Q$ by shifting the surrounding holes iteratively [38]. This will further suppress the excess radiation loss caused by the sharp transitions at cavity edges. A spatial resolution of 21.5 nm is used in combination with subpixel averaging. The starting optical $Q$ for nanobeams with ($w_{as}$, $w_b$, $L_s$) = (60 nm, 80 nm, 1.8$a$) is around 2.0 × 10$^3$. As shown in Fig. 4(a), shifting the adjacent holes towards the nanobeams with $s_{x1}$ = −0.18$a$ increases the optical $Q$ to 7.3 × 10$^3$. Further tuning of $s_{x2}$, $s_{x3}$, and $s_y$ leads to an optical $Q$ of 1.95 × 10$^4$. This optimized geometry with hole offsets ($s_{x1}$, $s_{x2}$, $s_{x3}$, $s_y$) = (−0.18$a$, −0.06$a$, 0.22$a$, −0.15$a$) will be subsequently referred to as "Design 1." It is worth noting that $s_y$ is not trivial in this L3-nanobeam cavity design: a variation of $s_y$ from 0 to −0.15$a$ actually doubles the optical $Q$. Experimental optical $Q$ values higher than 10$^4$ are measured based on Design 1 [32]. Following a similar optimization process, cavities with thinner nanobeams ($w_{as}$, $w_b$, $L_s$) = (60 nm, 60 nm, 1.9$a$) exhibit an even higher optical $Q$ of 5.22 × 10$^4$ with hole offsets ($s_{x1}$, $s_{x2}$, $s_{x3}$, $s_y$) = (−0.3$a$, −0.02$a$, 0.1$a$, 0), which will be subsequently referred to as "Design 2." These two designs with different beam geometries demonstrate the achievable high optical $Q$ of such L3-nanobeam cavities.

To further investigate the cavity mode, the modal distributions of the $E_y$ component for different $Q$ values are analyzed using spatial Fourier transform (FT) [38] as shown in Fig. 5. The fields have an odd symmetry in the $x$ direction, leading to a negligible portion inside the radiation light cone as indicated by the red circles [45]. Spatial components at $K_x \approx 0$ of the leaky fields are thus greatly suppressed. From Fig. 5(d) to (f), the leaky components inside the light cone are reduced as the optical $Q$ increases. This indicates that the enhanced optical $Q$ during the optimization process results not only from a smoother field profile but also from a more delocalized field [see Fig. 5(a)–(c)]. The optical modal volumes for the two designs in Fig. 5(b) and (c) are 0.079 $\mu$m$^3$ and 0.12 $\mu$m$^3$, or, 0.021 ($\lambda_0/n_{air}$)$^3$ and 0.032 ($\lambda_0/n_{air}$)$^3$, respectively, where $\lambda_0$ is the free-space resonant wavelength and $n_{air}$ is the refractive index of air. The small optical modal volumes compared to other high-$Q$ PhC cavities [38] are actually a result of predominantly localized modal energy inside the air slots. So we have demonstrated the high optical $Q$ and small modal volumes in L3-nanobeam cavities with flexible designs of embedded nanobeams.

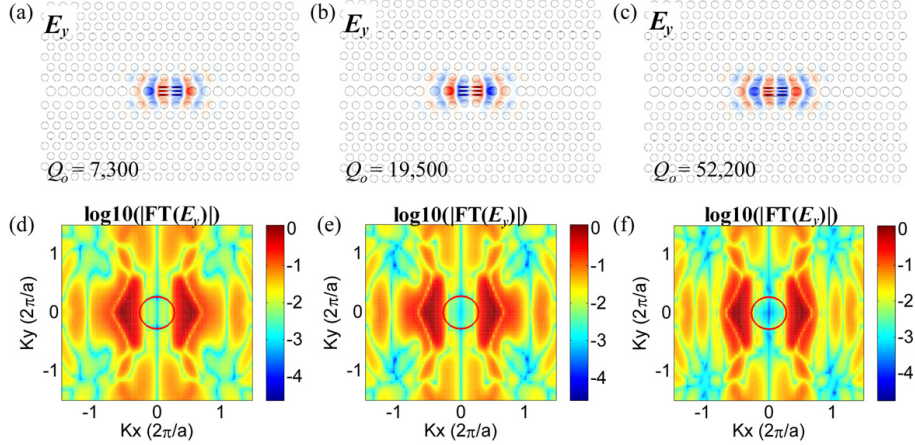

Fig. 5. (a)–(c) Modal distribution of the $E_y$ field. (d)–(f) the corresponding spatial Fourier transformation (FT) for cavities with increasing optical $Q$. (a) and (d) correspond to $(L_s, s_{x1}) = (1.8a, -0.18a)$ in Fig. 4(a) with an optical $Q$ of $7.3 \times 10^3$. (b) and (e) correspond to the optimized geometry Design 1. (c) and (f) correspond to the optimized geometry Design 2.

## 3. Mechanical design: eigenmodes and elastic radiation leakage

Doubly clamped beams have been widely used to build mechanical resonators [35–37]. Their high mechanical quality factor $Q_m$ and mechanical frequency $f_m$ are very useful for optomechanical applications, such as mass sensing, force sensing, and cooling/amplification of mechanical vibrations [46]. Depending on specific geometry, the frequency of the silicon nanobeams used here is around 1 GHz for the fundamental in-plane mode. The mechanical $Q$ is affected by various loss mechanisms, e.g., clamping, thermoplastic damping, defect motions, and fluidic loss, etc. Among all the sources, clamping loss is usually a major loss channel for doubly clamped beams [37]. An intuitive picture is that the vibrating nanobeams apply a force to the anchors, exciting elastic waves that carry away the mechanical energy to the environment. A challenge in modeling is that the computation domain always has a finite size. Thus, nonreflective boundaries should be applied to mimic an open system. The easiest implementation is to place some lossy material that absorbs the outgoing waves [47]. However, spurious reflections arise easily from the sudden change of material impedance. Additionally, because of the low absorption efficiencies for waves of arbitrary incident angles, large absorbing pads are usually required, which takes considerable amount of computation resources. Here, a perfectly matched layer (PML) is used for the mechanical simulation. The PML can be viewed as a very efficient impedance-matched nonphysical material [33,34]. Its implementation is based on complex coordinate scaling and has also been widely used in electromagnetic simulations [44]. The mechanical $Q$ can be obtained in various ways: from e.g., driven response, resonant amplitude decay, and complex resonant frequency. In finite-element simulation, the mechanical $Q$ is usually determined from the complex resonant frequency method where $Q_m = \text{Re}(f_m)/2\text{Im}(f_m)$, with $\text{Re}(f_m)$ and $\text{Im}(f_m)$ being the real and imaginary parts of the eigenfrequency $f_m$, respectively. The methods based on driven response or resonant amplitude decay are much more computationally costly.

The complex frequencies of mechanical modes of the L3-nanobeam cavity are calculated with the eigenvalue module in COMSOL 4.2a, a multiphysics solver based on a finite-element method [48]. As shown in Fig. 6(a), the simulated geometry is identical to that used in the optical modeling. PMLs are applied at the boundaries of computation domain for absorbing the radiating elastic waves without reflection. The outer edges of the PMLs are fixed, and the thickness of the PMLs is set approximately to one elastic wavelength. As shown in Fig. 6(a), the minimum mesh element size is 2.4 nm and the mesh element growth rate is 1.3. The maximum mesh element size is 20 nm in the beam region and 60 nm in other regions. The resolution of the curvature is 0.2. With these settings, every mesh element is thus at least 89 times smaller than the wavelength of the transverse and longitudinal elastic waves (9.66 μm and 5.34 μm, respectively, in silicon at 1 GHz). The frequency lower, the wavelength gets longer and the relative resolution gets even finer. The effectiveness of absorption by the PMLs is proved in Ref. [34] by applying harmonic point forces to a membrane.

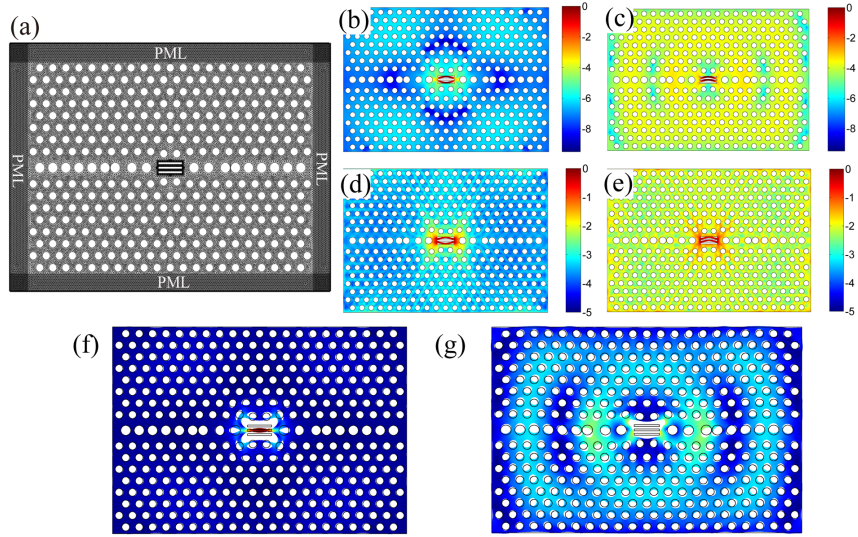

Fig. 6. (a) Top view of the meshed structure used in the finite-element analysis. Fixed boundary conditions are applied outside the PMLs. The top and bottom surfaces are set as free boundaries. (b),(c) Normalized displacement field intensity (log scale) for the differential and common mode of Design 1, respectively. (d),(e) Corresponding von Mises stress field (log scale). (f) Radiating longitudinal elastic wave excited by the differential beam motion. (g) Radiating transverse wave excited by the common beam motion. In (f) and (g), the displacement fields (linear scale) are overlaid with structural deformation (movies attached separately).

Next, the properties of the in-plane mechanical modes of the nanobeams are investigated. The frequency $f_m$ of the fundamental mode of doubly clamped beams is expressed as $f_m = C\sqrt{E/\rho}\,(w_b/L_s^2)$ [37], where the Young's modulus $E$ is 170 GPa, and the density $\rho$ is 2330 kg/m$^3$ for single-crystal silicon. $C$ is a constant dependent on the mode and the beam clamping conditions at the ends of the beam. For the fundamental mode of a doubly clamped beam, $C =$ 1.07 (for a Poisson's ratio of 0.28) and the estimated frequency is 1.2 GHz for a beam with the dimensions of Design 1. For the L3-nanobeam cavity, the two nanobeams are clamped to the PhC slab on either side and are thus mechanically coupled via the anchors, resulting in a differential and a common mode [30]. For Design 1, the complex frequencies of these eigenmodes are $9.61 \times 10^8 + i4.24 \times 10^4$ Hz and $9.58 \times 10^8 + i9.94 \times 10^6$ Hz, respectively as determined from the finite-element simulations. These frequencies are lower than the value calculated above due to the finite mechanical compliance at the clamping points. On the other side, the clamping-loss-limited mechanical $Q$ for the differential mode ($1.13 \times 10^4$), is more than two orders of magnitude higher than that for the common mode (only 48). This significant $Q$ difference can be explained with the mechanical displacement field and stress field shown in Fig. 6(b)–(g). Fig. 6(b) and (c) are snapshots of the mechanical displacement intensity (log scale) for the differential and common modes. The displacement intensity is defined as $I = \sqrt{\mathrm{Re}(u)^2 + \mathrm{Re}(v)^2 + \mathrm{Re}(w)^2}$ where $u$, $v$, and $w$ are displacements in the $x$, $y$ and $z$ directions, respectively. The difference resides not only in the beam motion but also in the radiating elastic waves propagating in the PhC membrane: First, the amplitude of the radiating elastic waves for the differential mode is much smaller than that for the common mode, which directly explains the much higher $Q$ of the differential mode than the common mode. Second, the radiation pattern for the common mode is similar to that produced by an in-plane harmonic point force driving the membrane in the $y$ direction, while the radiation pattern for the differential mode is more interference-like. The loss channels are also different for these two modes, as illustrated by the stress fields in Fig. 6(d) and (e). The differential mode has much more localized field around the anchor region than the common mode. This is attributed to the fact that the two nanobeams pull the anchors in opposite directions for the differential mode, whereas they pull in the same direction for the common mode. The different forces induce the excited elastic waves with different phases for the two modes. Further inspection of the propagating elastic waves reveals their different nature. As shown in Fig. 6(f) and (g) (movies attached separately), the lattice holes move along the propagation direction for the differential mode, but move laterally with respect to the propagation direction for the common mode. This indicates that longitudinal waves are excited by the differential beam motion while transverse waves are excited by the common beam motion, also in agreement with their different stress fields. The above investigation points to a greatly suppressed radiation loss for the differential mode. Experimental results [32] based on Design 1 have shown mechanical $Q$ values up to 1230 measured in vacuum, where the variation is mainly caused by fabrication imperfections, which will be discussed subsequently.

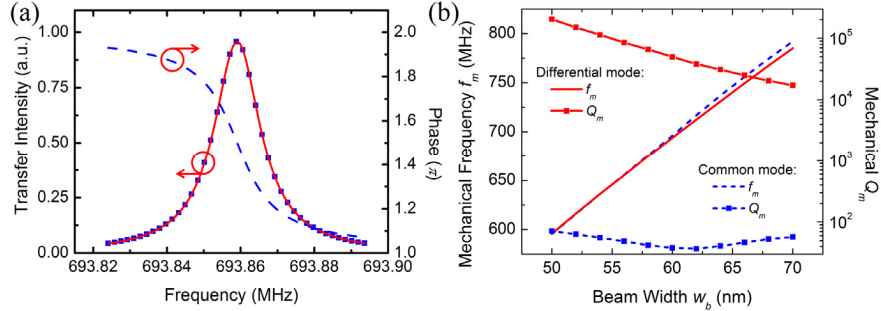

Fig. 7 (a) Transfer function obtained by forced frequency–response analysis. The phase, the transfer intensity, and its Lorentzian fit are plotted in blue dashed line, blue square markers, and red solid line, respectively. (b) Mechanical frequency $f_m$ and quality factor $Q_m$ versus the beam width for both the differential and common mode. Design 2 with $w_b = 60$ nm is used here.

Compared with Design 1, Design 2 employs thinner and longer nanobeams ($w_b$, $L_s$) = (60 nm, 1.9$a$). The simulated frequency is 693.9 MHz for the differential mechanical mode, about 268 MHz lower than that of Design 1. The mechanical $Q$ calculated from the complex frequency is 4.99 × $10^4$, more than four times higher than that of Design 1. Its transfer function between the force and displacement obtained from the frequency–response module in COMSOL 4.2a is shown in Fig. 7(a). A harmonic point force oscillating along the $y$ direction is applied at the center of one beam. The transfer intensity is the peak-normalized square of the displacement amplitude, from which a mechanical $Q$ of 4.53 × $10^4$ is obtained by Lorentzian fitting. The slight difference from the value obtained from the complex frequency originates from the computational errors between different solvers. The $\pi$-phase transition across the resonance also provides the mechanical $Q$ from its maximum phase slope $2Q_m/f_m$, which is essentially the same as the fitted value for transfer intensity. The transfer function method thus confirms the results obtained by the eigenvalue solver. However, since it is time-consuming to implement the frequency response calculation in large 3D modeling, the eigenvalue solver is usually preferred for the mechanical $Q$ analysis. The above numerical results from two different designs have shown a large frequency tuning range and a strong dependence of mechanical $Q$ on the beam width. Such dependence is exhibited in Fig. 7(b), where the beam width varies from 50 to 70 nm for Design 2. The mechanical frequencies change linearly with a slope of 9.45 MHz/nm and 9.91 MHz/nm for the differential and common mode, respectively. The slight difference between their slopes is a result of varying mechanical coupling strength of the two beams: wider beams with larger elastic constant exert larger forces to the clamping points yielding stronger coupling and larger frequency difference. The L3-nanobeam cavities possess high frequencies for beam width $w_b$ less than 100 nm with a linear frequency dependence on $w_b$ according to $f_m \propto w_b/L_s^2$. Furthermore, the mechanical $Q$ increases from $1.72 \times 10^4$ to $2.03 \times 10^5$ as the beam width decreases from 70 to 50 nm. This is most likely due to the facts that beams with a smaller elastic constant and a smaller mass apply less force to the anchor region and that the stress field is more localized with a reduced beam cross-sectional area. These two factors lead to weaker residual loss (and thus higher $Q$) for the differential modes, while they do not help much to the $Q$ for the common mode. It should be noted that, with the beam width variation, the optical $Q$ values remain above $4.36 \times 10^4$ as shown in Table 1. The overall resonant wavelength shift is about 19 nm, corresponding to an average of 0.95 nm per 1-nm increase in beam width.

## 4. Dispersive optomechanical coupling and influence of beam asymmetry

In the L3-nanobeam cavities, the optical and mechanical modes are mutually coupled. On one hand, optical forces created by the injected photons inside the cavity modify the static positions and dynamic response of the mechanical beams. On the other hand, the beams' motion changes the phase of the optical cavity field, inducing a shift of the resonant optical wavelength. Both of these effects are directly related to an optomechanical coupling rate, which characterizes the strength of optomechanical transductions [1,37]. The dispersive optomechanical coupling rate is defined as $g_{om} = d\omega_o/dx$ where $\omega_o$ is the angular frequency of the optical mode and $x$ represents the amplitude of the mechanical motion. The corresponding vacuum optomechanical coupling rate is defined as $g_0 = g_{om}x_{zm}$, where $x_{zm}$ is the zero-point motion of the mechanical mode given by $x_{zm} = \sqrt{\hbar/2m_{eff}\Omega_m}$ with $m_{eff}$ the effective modal mass and $\Omega_m$ the angular mechanical frequency ($2\pi f_m$). For regular cavities like microtoroids, micro-disks, and Fabry–Pérot etalons, $g_{om}$ is easily determined from their characteristic lengths [1]. For cavities with complicated geometries like our L3-nanobeam cavities, $g_{om}$ has to be numerically calculated based on the unperturbed optical and mechanical fields by using first-order perturbed solutions of Maxwell's equations with shifting boundaries [49,50]. Ref. [30] provides an expression for this with consistent definitions of effective modal volume $V_m$ and mass $m_{eff}$. Table 1 lists all the numerical results, where two features are worth noting: For different designs, the common mode always has slightly larger $V_m$ and $m_{eff}$ than the differential mode due to the mechanical mode delocalization as previously discussed. The $g_{om}/2\pi$ for the common mode is zero, since, to first order, the effective refractive index of the cavity region does not change with the in-plane beams' motion as dictated by symmetry. Properties for the two extreme structures based on Design 2 in Fig. 7(b) are also included in Table 1. The structure with $w_b = 50$ nm possesses the smallest effective mass, while the one with $w_b = 70$ nm achieves the maximum optomechanical coupling rate. Due to their high optomechanical coupling and high mechanical $Q$, the differential modes can easily be detected by optical transduction in experiments [27,30,32].

So far we have been focusing on symmetric structures with two identical beams. As both nanobeam width and slot width are below 100 nm, tiny variations from the ideal design can cause deviation from the expected properties. To evaluate such effects from, e.g., fabrication imperfections, the center slot of Design 1 is shifted laterally such that the two nanobeams now have different beam widths. Fig. 8(a) shows that the higher frequency mode (Mode 1) originates from the differential motion of the two beams and thus exhibits a higher mechanical $Q$ than the lower frequency mode (Mode 2) until the center slot shift $s_c$ reaches 2.0 nm. Note that the mechanical $Q$ of Mode 1 drops to around 1,000 with $s_c = 0.25$ nm, i.e., only a 0.5-nm difference in the beam width. It continues to drop to around 100, similarly to that of Mode 2 when $s_c$ is larger than 1.0 nm. The decoupling of the two beams with increased $s_c$ is also reflected by the beam frequencies, both of which exhibit a linear dependence for $s_c$ larger than 0.5 nm, each approaching the frequency of an individual beam. When $s_c$ reaches 3.0 nm, the frequency difference between the two branches is as large as 50.8 MHz. Fig. 8(b)–(e) show the displacement fields of the nanobeams for the two modes with $s_c = 0.25$ nm (b,c) and $s_c = 3.0$ nm (d,e). The coupled and uncoupled beam motions are evident, which is consistent with the frequency behavior shown in Fig. 8(a). It should be noted that such a 3-nm lateral shift of center slot only slight changes the optical $Q$, but considerably alters the optomechanical coupling rates for both mechanical modes to approximately one half of that of the differential mode of Design 1 [see Table 1].

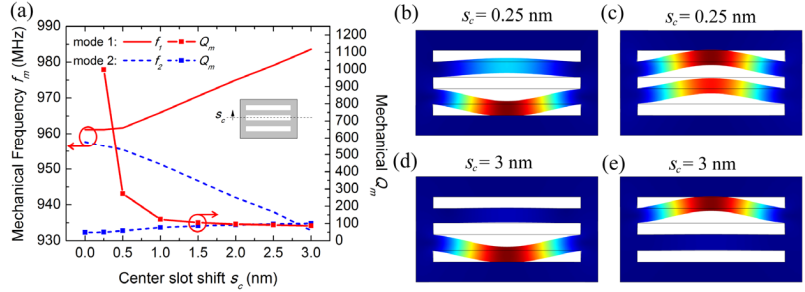

Fig. 8. (a) Mechanical frequency $f_m$ and quality factor $Q_m$ versus the center-slot displacement $s_c$. Design 1 corresponds to the structure with $s_c = 0$. Mode 1 originates from the differential mode, while Mode 2 originates from the common mode. (b)–(e) Zooms of the displacement fields of the nanobeams with $s_c = 0.25$ nm (b,c) and $s_c = 3.0$ nm (d,e).

Table 1. Optomechanical properties of the L3-nanobeam cavities. Design 1 and Design 2 refer to the two structures obtained in optical $Q$ optimization for different beam geometries.

| Structure | $\lambda_0$ (nm) | $Q_o$ | Mode | $f_m$ (MHz) | $Q_m$ | $V_m$ ($\mu m^3$) | $m_{eff}$ (fg) | $g_{om}/2\pi$ (GHz/nm) | $g_0/2\pi$ (kHz) |
|---|---|---|---|---|---|---|---|---|---|
| Design 1 | 1541.7 | $1.95 \times 10^4$ | Diff. | 961.2 | $1.13 \times 10^4$ | 0.011 | 26.5 | 11.3 | 204.6 |
| | | | Com. | 957.7 | 48 | 0.013 | 29.5 | ~0 | ~0 |
| Design 2 | 1553.2 | $5.22 \times 10^4$ | Diff. | 693.9 | $4.99 \times 10^4$ | 0.0087 | 20.4 | 10.9 | 265.4 |
| | | | Com. | 696.1 | 38 | 0.0093 | 21.6 | ~0 | ~0 |
| Fig. 7(b), $w_b = 50$ nm | 1543.1 | $4.36 \times 10^4$ | Diff. | 595.9 | $2.03 \times 10^5$ | 0.0072 | 16.8 | 6.8 | 197.6 |
| | | | Com. | 595.5 | 71 | 0.0074 | 17.2 | ~0 | ~0 |
| Fig. 7(b), $w_b = 70$ nm | 1562.2 | $5.32 \times 10^4$ | Diff. | 785.0 | $1.72 \times 10^4$ | 0.0103 | 24.1 | 15.5 | 326.6 |
| | | | Com. | 792.0 | 57 | 0.0112 | 26.2 | ~0 | ~0 |
| Fig. 8(a), $s_c = 3$ nm | 1540.8 | $1.94 \times 10^4$ | 1 | 983.6 | 87 | 0.0063 | 14.7 | 6.1 | 145.6 |
| | | | 2 | 932.8 | 101 | 0.0058 | 13.4 | 4.8 | 124.1 |

## 5. Conclusions

We have numerically studied a novel L3-nanobeam cavity, which consists of two mechanical nanobeams embedded in a PhC membrane where three holes are removed. 3D modeling with PMLs is employed for both optical and mechanical simulations. With *ab initio* calculation and comprehensive optimization, an optical $Q$ up to $5.2 \times 10^4$ is obtained. The fundamental in-plane mechanical modes of the high-optical-$Q$ designs are also analyzed systematically. The mechanical frequencies approach 1 GHz and can easily be tuned by slightly varying the beam width while maintaining a high optical $Q$. The anchor-loss-limited mechanical $Q$ for the differential mode is higher than $10^4$. The elastic radiation waves are shown to be transverse for the common mode and longitudinal for the differential mode. The effects on mode decoupling and mechanical $Q$ due to the fabrication imperfections are also studied. The optical transduction efficiency of the differential mode is very high with vacuum optomechanical coupling rates over 200 kHz. Such femtogram mass, high optical $Q$, high mechanical $Q$ structures are promising for optomechanical applications, especially in the ultrasensitive measurements involving mass, force, and displacement. In principle, an ultrahigh-$Q$ optical cavity can also be designed based on the mode-gap effect [51] by engineering the slotted waveguide band shown in Fig. 2. However, the mechanical frequency is expected to be much lower because of the much longer beams. Conceptually, the L3-nanobeam cavities represent a new type of nano-optomechanical systems created by directly placing ultrasmall mechanical resonators into a photonic nanocavity.

## Acknowledgments

The authors thank Jie Gao and James F. McMillan for helpful discussions on the optical and mechanical simulations. This work is supported by Defense Advanced Research Projects Agency (DAPPA) DSO with program manager Dr. J. R. Abo-Shaeer under contract No. C11L10831. M.P. acknowledges a Rubicon fellowship from The Netherlands Organization for Scientific Research (NOW)/Marie Curie Cofund Action.